\documentstyle[aps,prd,preprint,tighten]{revtex} 
\begin{document} 
\draft
\preprint{IMSc./2000/08/46}
\title{3+1 dimensional Yang-Mills theory as a local theory of evolution of metrics
 on 3 manifolds.}
\author{Pushan Majumdar \thanks{e-mail:pushan@imsc.ernet.in} \and
\addtocounter{footnote}{1}
H.S.Sharatchandra \thanks{e-mail:sharat@imsc.ernet.in}}
\address{Institute of Mathematical Sciences,C.I.T campus Taramani. 
Madras 600-113} 
\maketitle
\begin{abstract} 
An explicit canonical transformation is constructed to relate the physical subspace 
of Yang-Mills theory to the phase space of the ADM variables of general relativity.
This maps 3+1 dimensional Yang-Mills theory to local evolution of metrics on 3 manifolds.
\end{abstract}
The question of whether the dynamics of Yang-Mills theory can be completely captured in 
terms of gauge invariant quantitites has been raised many times. This is important 
especially because of confinement in QCD. One approach has been to rewrite the theory 
as dynamics of the loop variables. These Wilson loops are non-local variables and they 
also form an over-complete set. The possibility of using the gauge invariant combination
${\vec E}^i\cdot{\vec E}^j$ of the non-Abelian electric field has also been explored
\cite{1}\cite{2}\cite{2a}.
In another approach, the gauge invariant variables ${\vec B}_i[A]\cdot{\vec B}_j[A]$
have also been considered \cite{4}\cite{5}.
Analogy to gravity yields a nice geometric interpretation for 2+1 dimensional Yang-Mills 
theory \cite{9}. Such an approach for 3+1 dimensions has been attempted in \cite{10}

In this article we use certain techniques motivated by the Ashtekar formulation of 
gravity \cite{7}. We map the physical phase space of Yang-Mills theory to the phase space of 
the ADM variables of general relativity by an explicit canonical 
transformation. To do this we augment the ADM variables ($g_{ij}, \pi^{ij}$) by a set 
of auxiliary variables ($\theta_a, \chi^a$) to match in number, the variables ($A_i^a, 
E^{ia}$)
of the extended phase space of Yang-Mills theory. It turns out that the non-Abelian 
Gauss law simply becomes the constraint $\chi^a=0$. Therefore the physical subspace
of the phase space of the Yang-Mills theory is exactly mapped to the phase space of
the ADM variables and the dynamics can be rewritten as a local theory of evolution of 
metrics on 3-manifolds.

We use the language of functional integrals, but every step below may be interpreted in 
terms of dynamics of the classical theory.
We begin with the Euclidean partition function
\begin{equation}\label{Ef}
Z=\int\:{\cal D} 
A_{\mu}^{a}\;exp\{-\frac{1}{4g^{2}}\int\:(\partial_{\mu}\vec{A}_{\nu}
-\partial_{\nu}\vec{A}_{\mu}+\vec{A}_{\mu}\times\vec{A}_{\nu})^2\}.
\end{equation}
Introducing an auxiliary field $E^{ia}$, and integrating over $A_0^a$,
we get
\begin{equation}\label{fgauss}
Z=\int\:{\cal D} A_{i}^{a}{\cal D} E^{ia}\;\delta (D_{i}[A]E^{i})\;
 exp \{ \int (-{\cal H}+i\vec{E}^{i}\cdot\partial_{0}\vec{A}_{i})\}
\end{equation}
where
\begin{equation}\label{ham}
{\cal H}=\frac{1}{2} (g^{2}E^{2}+\frac{1}{g^{2}}B^{2}),
\end{equation}
is the hamiltonian density, and
\begin{equation}
\vec{B}^{i}[A]=\frac{1}{2}\epsilon^{ijk}(
\partial_{j}\vec{A}_{k}-\partial_{k}\vec{A}_{j}+\vec{A}_{j}\times\vec{A}_{k})
\end{equation}
is the non-Abelian magnetic field. 
Using the Feynman time slicing procedure, it is also clear that $A_{i},E^{i}$ are the 
conjugate variables of the phase space.
There are also three first class constraints, the non-Abelian Gauss law :
\begin{equation}\label{gauss1}
D_{i}[A]\vec{E}^{i}=0.
\end{equation}

Motivated by the Ashtekar variables, we define a driebein $e$ by
\begin{equation}\label{2}
{\vec E}^i=\frac{1}{2}\epsilon^{ijk}{\vec e}_j\times {\vec e}_k
\end{equation}
Assuming $||E||=||e||^2$ is nonzero, we can invert (\ref{2}) to get 
$ e_i^{a} = ||E||^{\frac{1}{2}}(E^{-1})_i^{a}$.
Define ${\bar A}[E]$ as the connection one form which is torsion-free with 
respect to the driebein ${\vec e}_i$. We have 
\begin{equation}\label{3}
\epsilon_{ijk}(\partial_j {\vec e}_k + {\vec {\bar A}}_j(E)\times {\vec e}_k)=0
\end{equation}
Therefore 
\begin{equation}\label{3a}
D_{i}[{\bar A}(e)]\vec{E}^{i}=0
\end{equation}
is identically valid. Hence we may 
replace Gauss law (\ref{gauss1}) by 
\begin{equation}\label{4}
{\vec a}_i\times {\vec E}^i=0
\end{equation}
where $a_i= A_i-{\bar A}_i(E)$ transforms homogeneously under gauge transformation.

We now observe that the change of variables from $\{A_i, E^i\}$ to $\{a_i, E^i\}$
is a canonical transformation. Consider the generating function
\begin{equation}\label{5}
S[a_i^a, E^{ia}]=\int d^3x\;a_i^aE^{ia}\;+\;{\bar S}[e]
\end{equation}
where 
\begin{equation}\label{6}
{\bar S}[e]=\frac{1}{2}\int d^3x\, \epsilon^{ijk}{\vec e}_i\cdot\partial_j{\vec e}_k
\end{equation}
The momentum conjugate to the new coordinate $a_i^a$ is 
$\frac{\delta S}{\delta a_i^a}=E^{ia}$, same as for $A_i^a$. The relation between the 
old and the new coordinates is $A_i^a=\frac{\delta S}{\delta E^{ia}}$ so that 
$A_i^a-a_i^a=\frac{\delta {\bar S}}{\delta E^{ia}}$.
Now
\begin{equation}\label{7}
\frac{\delta {\bar S}[e]}{\delta e_l^d}=\epsilon^{ljm}\partial_je_m^d
=-\epsilon^{ljm}\epsilon^{def}{\bar A}_j^e[E]e_m^f.
\end{equation}
Therefore we have,
\begin{equation}\label{8}
\frac{\delta {\bar S}}{\delta E^{ia}}=\frac{\delta e_l^d}{\delta E^{ia}}
\frac{\delta {\bar S}[e]}{\delta e_l^d}={\bar A}_i^e[E]
\end{equation}
since
\begin{equation}\label{8a}
\frac{\delta e_l^d}{\delta E^{ia}}=\frac{1}{||e||}(\frac{1}{2}e_l^de_i^a-e_i^de_l^a).
\end{equation}

We now show that there exists a canonical transformation from the phase space $(a,E)$ 
with the constraint (\ref{4}) to the phase space of the ADM variables ($\pi^{ij}, g_{ij}$).
(The first class constraints involving the ADM variables, related to space-time 
translations is not relevant for the present context of Yang-Mills theory.) As the variables
$(a,E)$ are more in number than ($\pi, g$), we first augment the latter set by a canonically 
conjugate set $\{\chi^a, \theta_a\}$. Here
\begin{equation}\label{9} 
g_{ij}={\vec e}_i\cdot {\vec e}_j,
\end{equation}
and to define $\theta_a$ we consider the Polar decomposition of $e_i^a$:
\begin{equation}\label{10}
e_i^a=e_{ij}O^{ja}
\end{equation}
into a symmetric matrix $e_{ij}$ and an orthogonal matrix $O^{ja}$. $\theta^a$ is the lie 
algebra element corresponding to $O^{ja}$. 
\begin{equation}\label{11}
O=exp (i\theta_a \frac{T^a}{2}).
\end{equation}
The equations (\ref{9}-\ref{11}) give us the relation between the old momenta and the new 
coordinates. We relate the momenta via a generator of the canonical transformation
\begin{equation}\label{12}
S(\pi, \chi, E)=\int \left ( ({\vec e}_i\cdot {\vec e}_j)\pi^{ij}+ \theta_a (e)\chi^a 
\right )
\end{equation}
This gives us $g_{ij}\equiv\frac{\delta S}{\delta \pi^{ij}}={\vec e}_i\cdot {\vec e}_j$
as we want. Among the other quantitites, $\theta_a$ is given by
\begin{equation}\label{13}
\theta_a\equiv\frac{\delta S}{\delta \chi^{a}}=\theta_a[e].
\end{equation}
Also we have 
\begin{equation}\label{14}
a_i^a\equiv\frac{\delta S}{\delta E^{ia}}
\end{equation}
To express $a_i^a$ in terms of the other variables, we take the variation of $S$ 
with respect to $e_i^a$. We get
\begin{equation}\label{15}
\frac{\delta S}{\delta e_i^a}=2\pi^{ij}e_j^a + \frac{1}{2}\chi^b\epsilon_{ijk} 
M^{-1}[\theta]^b_c((e-{\bf 1}\,tr\,e)^{-1})_{kl}\,O^{ja}O^{lc}.
 \end{equation}
where $M$ is defined in the appendix.
Now note that 
\begin{equation}\label{16}
\frac{\delta S}{\delta e_i^a}=\frac{\delta E^{bj}}{\delta e_i^a}a_j^b.
\end{equation}
This explicitly relates $a$ to the other variables. We now show that $\chi$ is related 
to the Gauss law.
Contracting $\frac{\delta S}{\delta e_i^a}$ by $e_i^c$, we get from (\ref{16})
\begin{equation}
e_i^c\frac{\delta S}{\delta e_i^a}=-E^{ia}a_i^c+E^{ib}a_{ib}\delta_{ac} \label{17} 
\end{equation}
For the part antisymmetric in $(a,c)$, we get from (\ref{15}),
\begin{equation}\label{18}
({\vec a}_i \times {\vec E}^i)^a=\frac{1}{2}(M^{-1})_b^a[\theta]\chi^b
\end{equation}
Thus the canonical momentum $\pi^{ij}$ drops out in the Gauss law equation (\ref{gauss1}).
{\em With the new variables, the Gauss law is implemented by simply setting $\chi=0$.}

The partition function (\ref{fgauss}) in the new variables is ,
\begin{equation}\label{20}
Z=\int{\cal D}g_{ij}\,{\cal D}\pi^{ij}\,{\cal D}\theta^a\,{\cal D}\chi^a
\delta((M^{-1})^a_b[\theta]\chi^b) exp\int(-{\cal H}^{'}+i\pi^{ij}\partial_0g_{ij}
+i\chi^a\partial_0\theta^a).
\end{equation}
$\theta^a$ represents the gauge degrees of freedom. We may adopt the Faddeev-Popov 
procedure to choose $\theta^a=0$. In this case $M^a_b[\theta]=\delta^a_b$, and 
\begin{equation}\label{21}
Z=\int{\cal D}g_{ij}\,{\cal D}\pi^{ij}\,exp\int(-{\cal H}^{'}[g,\pi]+i\pi^{ij}
\partial_0g_{ij})
\end{equation}
Thus the functional integral is rewritten in terms of the conjugate variables $(g_{ij}, 
\pi^{ij})$ which are gauge invariant. The new Hamiltonian ${\cal H}^{'}$ is obtained from 
(\ref{ham}), by the replacements
\begin{eqnarray}
E^{ia}&\rightarrow & \frac{1}{2}\epsilon^{ijk}\epsilon^{abc}e_{jb}e_{kc} \\
A_i^a&\rightarrow & {\bar A}_i^a[E]+a_i^a \\
a_i^a&\rightarrow & \frac{1}{||e||}(\pi^{jk}g_{jk}e_{ia}-2\pi^{jk}g_{ik}e_{ja}) 
\end{eqnarray}
where $e_{ia}$ is regarded as the symmetric square root of $g_{ij}$.
Thus the $({\vec E}^i)^2$ term in the Hamiltonian becomes $\frac{g^{ii}}{||g||}$, while 
\begin{equation}
B_i[A] = B_i[{\bar A}[E]]+\epsilon_{ijk}D_j[{\bar A}[E]]a_k+\frac{1}{2}
\epsilon_{ijk}(a_j\times a_k)
\end{equation}
 corrsponding to an expansion about a `background' gauge
field ${\bar A}[E]$. The $({\vec B}_i[A])^2$ can be completely written as {\em local} 
expressions in $g_{ij}$ and $\pi^{ij}$. For example
\begin{eqnarray}
B_i^a[{\bar A}[E]] &=& \frac{1}{4||e||}\epsilon_{ijk}\,\epsilon_{lmn} R_{jl}\,g_{km}\,e_n^a  \\
\epsilon_{ijk}(a_j\times a_k)\cdot\epsilon_{imn}(a_m\times a_n)&=& {\vec a}_j\cdot {\vec a}_m
\,{\vec a}_j\cdot{\vec a}_m-({\vec a}_j\cdot{\vec a}_j)^2 
\end{eqnarray}
Similarly
\begin{equation}
(D_i[{\bar A}[E]]e_j)^a=\Gamma_{ij}^k[g]e_k^a
\end{equation}
where $\Gamma_{ij}^k$ is the affine connection corresponding to the metric $g_{ij}$.
$D_k[{\bar A}[E]]$ can be replaced by the covariant derivative corresponding to the affine 
connection $\Gamma_{ij}^k[g]$ when acting on $g_{ij}$ or $\pi^{ij}$. This way, $({\vec 
B}_i[A])^2$ can be written as a local expression in $g_{ij}$ and $\pi^{ij}$.

We have thus mapped the physical phase space of Yang-Mills theory onto the phase space of 
the ADM variables and the dynamics is now a local evolution of the metrics on 3-manifolds.
This completes the program envisaged in \cite{1}.

The canonical transformation constructed here can be used to map all the ADM constraints of 
general relativity to certain constraints on the Yang-Mills phase space without requiring 
complexification of the gauge field. In fact Barbero's constraints \cite{6} are reproduced.

Here we have related the gauge invariant combination ${\vec E}^i\cdot{\vec E}^j$ to a 
metric on a 3-manifold. It is also possible to construct a metric from the vector potential 
$A_i^a$ and use it to rewrite the Yang-Mills dynamics. These two approaches are dual of each 
other. This will be addressed elsewhere.

\subsection{Appendix}
To evaluate $\frac{\delta \theta^b[e]}{\delta e_i^a}$, note that 
\begin{equation}\label{aa}
e_i^a+\delta e_i^a=(e_{ij}+\delta e_{ij})exp(i(\theta [e] + \delta \theta [e])^b \frac{T^b}{2}).
\end{equation}
This gives
\begin{equation}\label{19}
\delta e_i^a \, O^{ka}=\delta e_{ik}+ \epsilon^{dae} e_{ij}O^{jd}O^{ka}{\bar{\delta \theta}}^e
\end{equation}
where ${\bar {\delta \theta}}^e=M^f_e [e]\delta \theta^e$ and the matrix $M^f_e$ is given by
\begin{equation}
O^T(\theta )O(\theta + \delta \theta )\approx 1+T^a M^a_b(\theta)\delta \theta^b.
\end{equation}
Taking the antisymmetric part of (\ref{19}) we can solve for $\frac{\delta \theta^b[e]}{\delta e_i^a}$.
We get
\begin{equation}
\frac{\delta \theta^b[e]}{\delta e_i^a}=-\epsilon_{ijk} M^{-1}[\theta]^b_c((e-{\bf 
1}\,tr\,e)^{-1})_{kl}\,O^{ja}O^{lc}.
\end{equation}
When two or all eigenvalues of the symmetric matrix $e_{ij}$ are degenerate, so are 
those of $(e-{\bf 1}\,tr\,e)$. In this case, the variables $\theta [e]$ equation (\ref{aa})
are ill defined and more care is required to define the new variables.

\end{document}